\documentclass[aps,prl,preprint,showpacs,groupedaddress]{revtex4}
\input{psfig.sty}
\begin{document}

\preprint{MPIKGF 2001-11}

\title{
Temperature- and Force-Induced $\beta$-Sheet Unfolding in an Exactly 
Solvable Model 
}

\author{Haijun Zhou}

\affiliation{Max-Planck-Institute of Colloids and Interfaces,
D-14424 Potsdam, Germany}

\date{\today}

\begin{abstract}
The stability of a $\beta$-sheeted conformation
and its transition into a random coil are studied with a 
2D lattice biopolymer model. At low temperature and low external
force, the polymer folds back and forth on itself and forms
a $\beta$-sheet. Our analytical calculation and Monte Carlo simulation
reveal that a co-operative $\beta$-sheet--random coil
transition takes places when the temperature or force is increased, 
with a dramatic decrease in the contact number. 
These predictions are in good agreement 
with experiments on titin protein. This transition is 
not a real phase-transition,
indicating that backbone hydrogen-bonding
alone is unable to stabilize a distinct $\beta$-sheet phase.
\end{abstract}

\pacs{87.15.-v,64.60.Cn, 05.50.+q, 82.35.Jk}

\maketitle
Single-molecule manipulation techniques were used by
Gaub and coworkers and several other  groups  
\cite{rief1997,carrion-vazquez1999} 
to study the unfolding  of 
proteins in real time. A  co-operative $``$all-or-nothing'' transition
between globular and randomly coiled states was observed at
a pulling force of $\sim 10^2$ piconewtons. 
A stretched coil segment
can be modelled as a wormlike chain of persistence length $\sim 10^{-1}$ 
nanometers \cite{rief1997,carrion-vazquez1999}, and 
several (atomic-level) molecular dynamics (MD) 
simulations have studied the
initial stage of unfolding and the final stage of refolding
(see, e.g., \cite{lu2000,paci1999}).
Monte Carlo (MC) simulations \cite{socci1999}
of lattice models also have been used to study the unfolding kinetics. 
The protein used in 
the experiments, namely
titin imunoglobulin  domains, is rich in $\beta$-sheets
\cite{rief1997,carrion-vazquez1999},  and 
numerical work demonstrated that the transition to a random coil
was initiated by the rupture of only a few hydrogen bonds between
adjacent $\beta$-strands \cite{lu2000}. Theoretically, however, 
it is still unclear  why the globule-coil transition
is a highly co-operative process, and  why there are 
dramatic differences between the unfolding behavior of helix-dominated
and $\beta$-sheet-dominated proteins \cite{rief1999}.

An issue of current interest is to study the 
generic statistical mechanics of biopolymer $\beta$-sheet
states. In this paper, the deformation and unfolding of
$\beta$-sheets are addressed within an exactly-solvable 2D model.
Because of the attractive interaction between monomers
along the chain, the polymer could fold back and forth on
itself, thereby  maximizing its contact energy.
However, a $\beta$-sheet, which is a relatively 
ordered structure and is favorable in energy, is deficit in configurational
entropy compared with an irregular coil configuration.
Because of this competition between energy and entropy, it is demonstrated
by our analytical and simulational calculations that, when the temperature or
the external force approaches some critical value, a small variation in
temperature or force could lead to a dramatic change in the polymer's
microscopic configuration. This  behavior corresponds well both with 
experimental observations and MD simulations on titin. 
However, we suggest that this
highly co-operative transition is not a phase-transition in the
strict sense; the nonspecific hydrogen-bonding interactions between 
the backbone molecules of a protein alone are hence unlikely to be able to
stabilize a distinct $\beta$-sheet phase.

\begin{figure}
\psfig{file=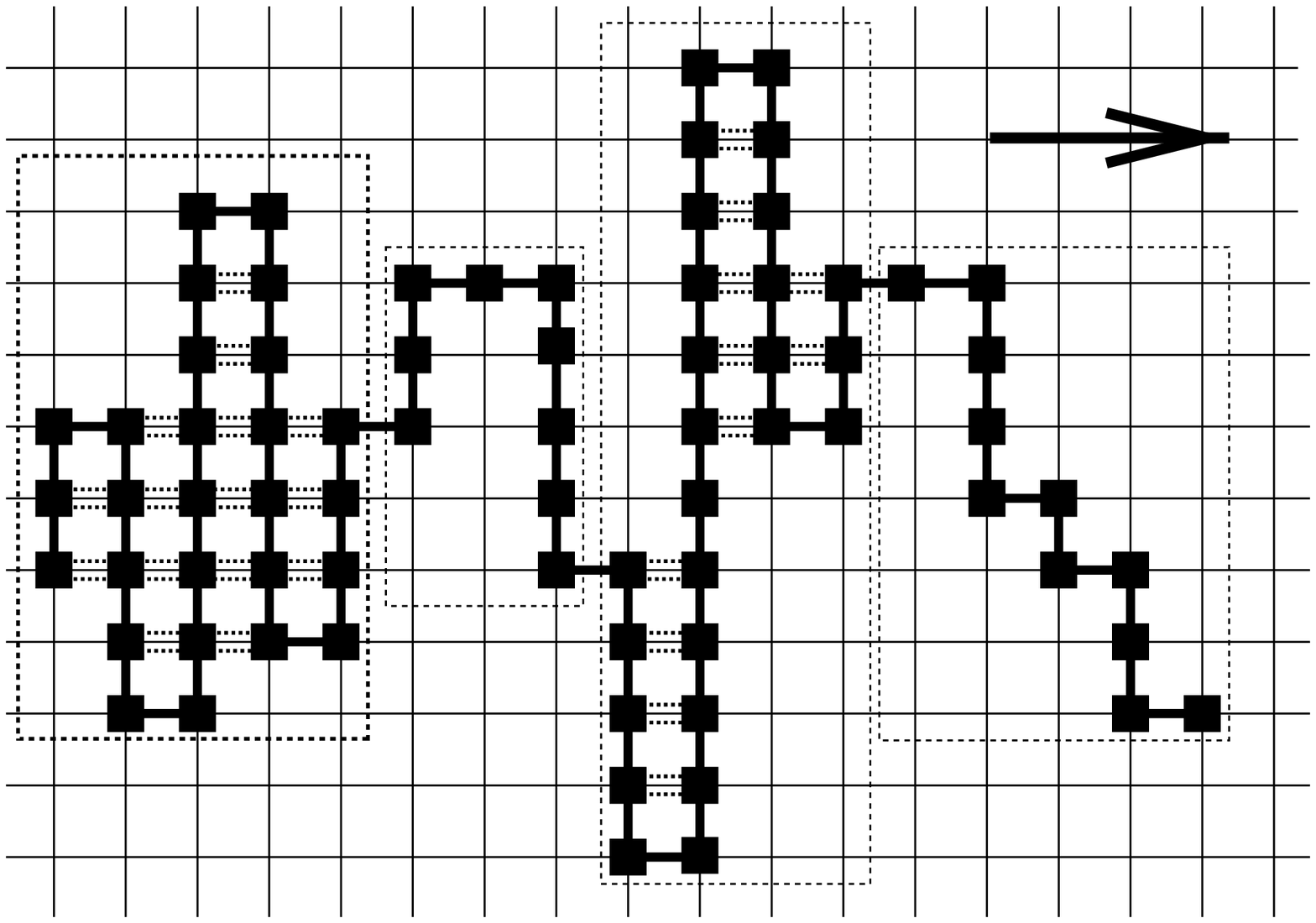,width=10.0cm}
\caption{\label{fig:fig01} A model directed biopolymer. Black squares 
are monomers and the  arrow denotes the ${\bf z}_0$ 
direction. The configuration shown here has two $\beta$-sheets and 
two coils.
}
\end{figure}

The mathematical model is constructed as follows:
Consider a biopolymer of $N$ identical units
(monomers) located on a square lattice (Fig.~\ref{fig:fig01}). 
The distance between two consecutive monomers is
$a_0$, and to  facilitate the calculation
the chain is assumed to be directed, namely the bond between monomers
$i$ and $i+1$ can not point to the $-{\bf z}_0$ direction  
(Fig.~\ref{fig:fig01}).
 If two monomers with indices $i$ and $i+m$ ($m\geq 3$) occupy nearest 
neighbor lattice sites,  an attractive energy of strength
$\epsilon$ is gained and these two monomers are then 
$``$in contact''. Each monomer can be in one of three states, 
with respectively zero, one, or two contacts. 
By folding back and forth on itself, the
polymer can gain contact energy, and a segment of
$m$ columns in which there is at least one pair of monomers in
contact between any two adjacent columns is referred to as a 
$``$$\beta$-sheet''. Two $\beta$-sheet
segments are separated by a $``$coil segment'' of $m$ columns ($m\geq 0$), 
in which all the monomers are free of contacts (Fig.~\ref{fig:fig01}).

We first discuss the free energy of  the $\beta$-sheet state.
Under the action of a force $f$ along direction ${\bf z}_0$, the 
energy of a $\beta$-sheet of $m$ columns is
$-\epsilon \sum_{i=1}^{m-1} {\rm v}(l_i,l_{i+1})- m f a_0$, where 
${\rm v}(l_i,l_{i+1})=l_i-1$ (if $l_i \leq l_{i+1}$) and 
$l_{i+1}-1$  otherwise;
$l_i$ is the number of monomers in the $i$-th column. The partition
function (PF) $Z_{\beta}(n)$ of a $\beta$-sheet of length $n$ 
is
\begin{equation}
Z_{\beta}(n)=\sum\limits_{m=2}^{[n/2]}\sum\limits_{l_1=2}\cdots
\sum\limits_{l_m=2} \delta_{l_1+\cdots+l_m}^n b^m 
\prod\limits_{i=1}^{m-1} a^{{\rm v}(l_i,l_{i+1})},
\end{equation}
where $\delta_i^j=1$ ($i=j$) and $0$  (otherwise); $a=\exp(\beta \epsilon)$,
$b=\exp(\beta f a_0)$, and $\beta=1/T$, $T$ being the temperature (the 
Boltzmann constant is set to $1$).

There can be many-body interactions among the monomers in a
$\beta$-sheet and a contact pair can be formed between two
remote monomers along the chain, consequently,   
direct evaluation of $Z_{\beta}(n)$ for a large system is difficult. 
To circumvent this, we  define an auxiliary  function
$
T_m(\zeta, k) = \sum_{l_1=2} \cdots \sum_{l_m=2}
\zeta^{k l_1} a^{{\rm v} (l_1,l_2)-l_1}
\prod_{i=2}^{m-1} (\zeta/a)^{l_i} a^{{\rm v}(l_i,l_{i+1})}
(\zeta/a)^{l_m}
$
for $m\geq 2$ and $T_1(\zeta,k)=\zeta^{2k}/a(a-\zeta^k)$. 
One finds  
\begin{equation}
T_m(\zeta,k)={{\zeta^{2k} T_{m-1}(\zeta,1)}\over {a (1-\zeta^k)}}
+{{\zeta^k (1-a)  T_{m-1}(\zeta,k+1)}\over {a (a-\zeta^k) (1-\zeta^k)}}.
\label{eq:T_m}
\end{equation}
Define $L_{\beta}(\zeta,k)=\sum_{m=1}^{\infty} b^m T_m(\zeta,k)$,
and from the recursive relation Eq.~(\ref{eq:T_m}) one obtains the
following two equations: 
(1)
$
L_{\beta}(\zeta,2)=A(\zeta) + B(\zeta) L_{\beta}(\zeta,1)
$
and (2)
$
L_{\beta}(\zeta,2)=H(\zeta)+J(\zeta) L_{\beta}(\zeta,1)
$.
After eliminating $L_{\beta}(\zeta,2)$ from these two equations,
one finds that 
\begin{equation}
L_{\beta}(\zeta,1)={A(\zeta)-H(\zeta) \over J(\zeta)-B(\zeta)}.
\label{eq:l_1}
\end{equation}
In Eq.~(\ref{eq:l_1}), $A(\zeta) = \zeta (1-\zeta)/ (a-1)$,
$B(\zeta)=(a-\zeta) (b\zeta^2+a\zeta-a)/b (a-1)\zeta$,
$H=(b/a)[\zeta^4 /(a-\zeta^2)+\sum_{i=3}^\infty \prod_{j=2}^{i-1}\phi(\zeta,j)
\zeta^{2 i}/(a-\zeta^i)]$, and
$J=(b/a)[\zeta^4/(1-\zeta^2)+\sum_{i=3}^\infty \prod_{j=2}^{i-1}\phi(\zeta,j)
\zeta^{2 i}/(1-\zeta^i)]$, where
$\phi(\zeta,j)=b(1-a){\zeta^k}/a (a-\zeta^k) (1-\zeta^k)$.

Both $H(\zeta)$ and $J(\zeta)$ are convergent when $\zeta\in [0, 1]$. When
$\Delta=1-\zeta\rightarrow 0^{+}$, our analytical and numerical
calculations suggest that $A(\zeta)-H(\zeta)\rightarrow \exp(-b/a\Delta)
f_{1}(\Delta)$ and $J(\zeta)-B(\zeta)\rightarrow
\exp(-b/a\Delta) f_{2}(\Delta)$, with $f_1(\Delta)/f_2(\Delta)
\rightarrow b/a(a-1)$.  Here $f_1(\Delta)$
and $f_2(\Delta)$ are two polynomials of $\Delta$ which contain
terms with negative powers of $\Delta$. 

\begin{figure}
\psfig{file=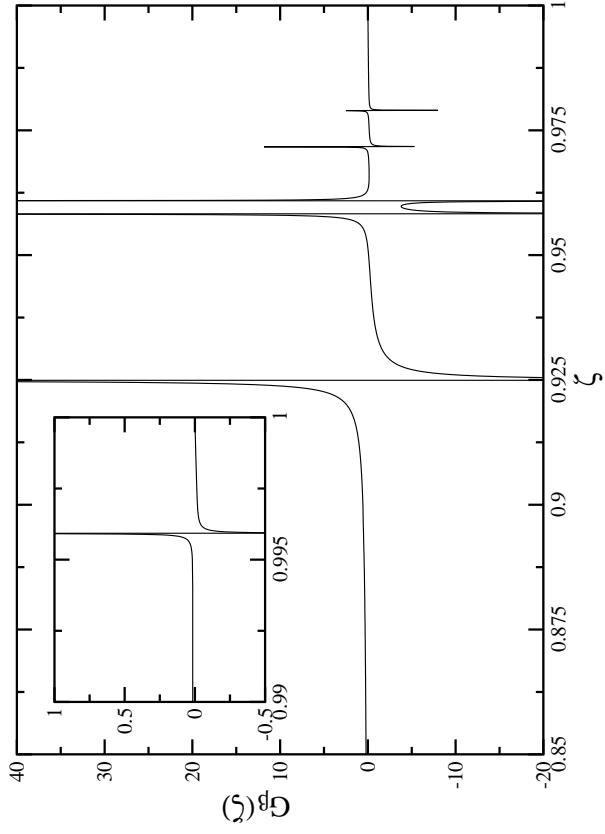,width=10.0cm}
\caption{\label{fig:fig02} The value of the partition function
$G_\beta(\zeta)$ as its divergent radius $\zeta_\beta$ is approached.
The temperature is set to $T=0.590928 \epsilon$ and the force is
$f=0.73866 \epsilon/a_0$ or $f=0$ (the inset).
}
\end{figure}

The grand partition function (GPF) of the $\beta$-sheet,
$G_{\beta}(\zeta)=\sum_{n=4}(\zeta/a)^n Z_{\beta}(n)$,
is nothing but $L_{\beta}(\zeta,1)-b^2 \zeta^2 /a(a-\zeta)$.
$G_{\beta}(\zeta)$ could be calculated based on Eq.~(\ref{eq:l_1})
at given $\zeta$. 
It keeps increasing as $\zeta$ approaches
$\zeta_\beta (<1)$ from $0$ and is divergent at $\zeta_\beta$, 
where $\zeta_\beta$ is the smallest positive root
of the equation $J(\zeta)-B(\zeta)=0$  (see Fig.
~\ref{fig:fig02}). The free energy density of a $\beta$-sheet 
segment equals to
$\rho_{\beta}=k_B T \ln (\zeta_\beta/a)$.  The divergence of $G_{\beta}$
at $\zeta_\beta$ indicates the absence of a distinct $\beta$-sheet phase.

Armed with the partition function of a $\beta$-sheet, 
we proceed to investigate its stability and the configurational
transition of the whole polymer system. 
To this end, one
needs to know the free energy of  a coil segment \cite{footnote1}. 
The total PF of a coil segment of length $n$ under the action of
an external force ($f$, incorporated in the
parameter $b$) is expressed as
\begin{equation}
Z_{\rm coil}(n)=2 \sum\limits_{m=1}^n 
Z_{\rm coil}^{\rm s}(n,m) b^m+ \sum\limits_{m=1}^n
Z_{\rm coil}^{\rm m}(n,m)b^m \label{eq:Z_coil}
\end{equation}
and $Z_{\rm coil}(0)=1$.
In this expression, 
$Z_{\rm coil}^{\rm s}(n,m)$ is  by definition
the total number of allowed
configurations for a  coil of length $n$ whose two ends are
separated by $m$ lattice steps along ${\bf z}_0$  and
whose last column has only a single monomer; 
$Z_{\rm coil}^{\rm m}(n,m)$
is defined similarly, but with the last column containing at least two
monomers.   
The coefficient $2$ in Eq.~(\ref{eq:Z_coil}) appears because 
the first column of a $\beta$-sheet  following 
a coil segment that ends with a column containing a single monomer
can have {\em two} possible orientations (Fig.~\ref{fig:fig01}).

One finds
$
Z_{\rm coil}^{\rm s} (n,m) = Z_{\rm coil}^{\rm s}(n-1,m-1)
+Z_{\rm coil}^{\rm m}(n-1,m-1),
$ 
and
$
Z_{\rm coil}^{\rm m}(n,m) = Z_{\rm coil}^{\rm s}(n,m+1)+ 
Z_{\rm coil}^{\rm s}(n-2,m-1),
$ 
with 
$Z_{\rm coil}^{\rm s}(0,m)=\delta_{m}^0$, $Z_{\rm coil}^{\rm
m}(0,m)=Z_{\rm coil}^{\rm m} (1, m)=0$,
$
Z_{\rm coil}^{\rm m}(2,m)=\delta_{m}^{1}$, and
$
Z_{\rm coil}^{\rm m}(n,1)=1-\delta_{n}^{1}$. 
With these preparations the GPF of the coiled state is
calculated to be
\begin{equation}
G_{\rm coil}(\zeta)=\sum\limits_{n=0}^{\infty}\zeta^n a^{-n} Z_{\rm coil}(n)
={{a (a-\zeta) (a+b \zeta)} \over {a^3-a^2 (b+1) \zeta -b^2 \zeta^3}},
\label{L:Coil}
\end{equation} 
It follows that the free energy density of a coil segment equals
$\rho_{\rm coil}=k_B T \ln(\zeta_{\rm coil}/a)$.
$\zeta_{\rm coil}$ is the point where $G_{\rm coil}(\zeta)=+\infty$:
$\zeta_{\rm coil}=\ln[(a/b)((\sqrt{\lambda}+b/2)^{1/3}-
(\sqrt{\lambda}-b/2)^{1/3})]$ with $\lambda=(b+1)^3/27+b^2/4$.

In the lattice model, any configuration of the polymer
is a chain of $\beta$-sheets and
random coils, with the two conformations occurring alternately
(Fig.~\ref{fig:fig01}). The total PF is then
\begin{eqnarray}
Z(N)&=&\sum\limits_{s=0}^{[N/4]}\sum\limits_{i_0=0}
\sum\limits_{j_1=4}\sum\limits_{i_1=0}\cdots\sum\limits_{j_s=4}
\sum\limits_{i_s=0}\sum\limits_{j_0=0} \nonumber \\
& &
\delta_{i_0+j_0+\sum_{k=1}^s (i_k +j_k)}^N 
\prod\limits_{k=0}^{s}
[Z_{\beta}(j_k)Z_{\rm coil}(i_k)], \label{Z:Total}
\end{eqnarray}
Finally the  GPF of the whole system  may be calculated:
\begin{equation}
G_{\rm total}(\zeta)=\sum\limits_{N=0}^\infty \zeta^N a^{-N} Z_{\rm total}(N)
={{[1+G_{\beta}(\zeta)]G_{\rm coil}(\zeta)}
\over {1-G_{\beta}(\zeta) G_{\rm coil}(\zeta)}}.
\label{eq:L_total}
\end{equation}

Because of the facts that $G_\beta(\zeta_\beta)=+\infty$ and 
$G_{\rm coil}(\zeta_{\rm coil})=+\infty$, it is always possible to
find a root ($\zeta_{\rm total}$) of the equation
\begin{equation}
G_{\rm coil}(\zeta_{\rm total}) G_{\beta}(\zeta_{\rm total})=1
\label{eq:zeta_total}
\end{equation}
in the range of $0<\zeta_{\rm total}<{\rm min}(\zeta_{\beta},
\zeta_{\rm coil})$.
Consequently, the free energy density of the whole system is 
neither that of the $\beta$-sheet state nor that of the
coil state: $\rho_{\rm total}=k_B T\ln(\zeta_{\rm total}/a)$.
There is no singularity in the total free energy expression, and hence no
real phase-transition process.
Almost forty years ago, it was revealed by intensified theoretical
investigations that 
the protein $\alpha$-helix--coil transition is not a phase-transition,
provided that only the  main-chain  hydrogen-bonding interactions
are considered
(see, e.g., \cite{lifson1964,poland1970});
we are surprised to find that the same conclusion holds for
the $\beta$-sheet transition.
Although there is no strict phase-transition between
$\beta$-sheet and random coil in the model,  our analytical and
simulational results presented below indicate that, when temperature
and the external force are low enough  the polymer is essentially in
the form of a $\beta$-sheet, and there is an rapid change in the 
polymer's total extension and its  total number of contacted pairs
when the temperature or the force is increased.  

\begin{figure}
\psfig{file=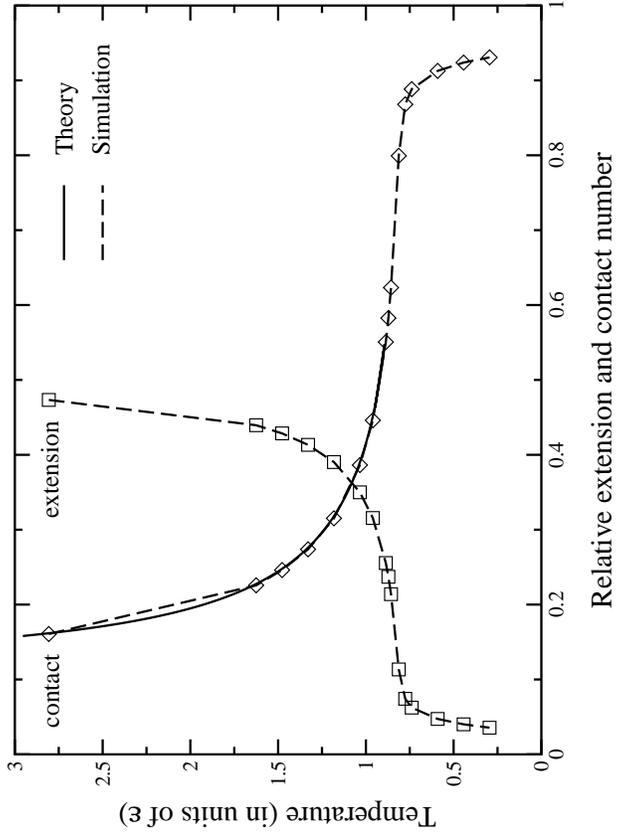,width=10.0cm}
\caption{\label{fig:fig03} 
Dependence of the contact number and the extension on the temperature
at zero force. Symbols are MC simulation data performed for
$N=1024$. 
}
\end{figure}

\begin{figure}
\psfig{file=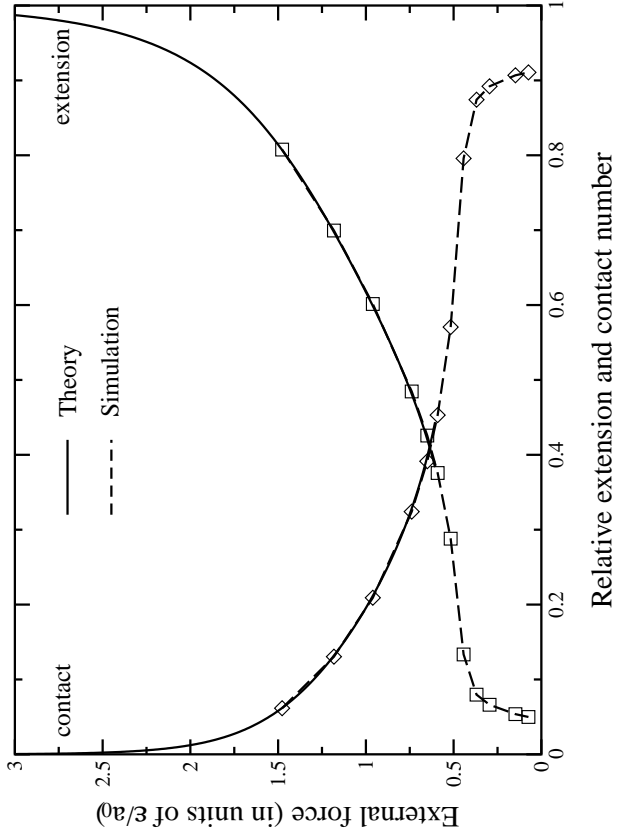,width=10.0cm}
\caption{\label{fig:fig04} Dependence of the contact number and the 
extension on the external force at fixed temperature $T=0.590928 \epsilon$.
Symbols are MC simulation data performed for $N=1024$.
}
\end{figure}

The total contact number $\sigma$ (scaled
by $N$) could be calculated by $\sigma=1-\partial
\ln (\zeta_{\rm total})/\partial \ln a$ and the total extension
$\ell$ (scaled by $N a_0$) could be calculated by $\ell=-b \partial 
\ln(\zeta_{\rm total})/\partial b$.
When there is no external force ($f=0$), the temperature--contact number
profile and the temperature--extension profile are shown in 
Fig.~\ref{fig:fig03}. And in Fig.~\ref{fig:fig04} the relationship
between force and extension and between force and contact number
are shown at a fixed temperature of $T=0.591 \epsilon$.
When the temperature and the force is
small, $\zeta_{\rm total}$ approaches $\zeta_\beta$ (the divergent radius
of $G_{\beta}$) and its value is difficult to be numerically determined
in much precision through Eq.~(\ref{eq:zeta_total}). Therefore, the 
analytical calculation in Fig.~\ref{fig:fig03} is shown only for 
$T\geq 0.8 \epsilon$ and that in Fig.~\ref{fig:fig04} for $f\geq 0.6 
\epsilon/a_0$. 
In Figs.~\ref{fig:fig03} and ~\ref{fig:fig04} the data obtained by
a MC simulation is also shown.
The Monte
Carlo simulation is performed with a polymer size of $N=1024$ (the 
results for $N=5000$ are essentially the same as the
data shown here). 
There are three kinds of  elementary updates of configurations:
(i) one monomer is moved  from one column to another column or stacked on the
right-hand side of the last column; (ii) the positions of two columns
are exchanged; (iii)
the direction of one column is changed by $180^\circ$. Special attention
is paid to make sure that  ergodicity and detailed balance are obeyed. Each
simulational data is obtained through $10^9$ MC steps (it takes about
$35$ hours in a Compaq station). The energetically 
most favorable state is a $\beta$-sheet whose
width and height both equal to $\sqrt{N}$. Therefore,
in the numerical simulation data shown in Figs.~\ref{fig:fig03} and
\ref{fig:fig04} for a chain of $N=1024$,
one has $\sigma\leq 0.935$
and $\ell\geq 0.0313$. For an infinite system,
$\sigma\rightarrow 1$ and $\ell\rightarrow 0$ as $T\rightarrow 0$.

Figure \ref{fig:fig03} indicates that when $f=0$ and $T<0.75 \epsilon$ almost
all the monomers of the polymer are in the $\beta$-sheet state. At $T\simeq
0.75 \epsilon$ there is a plateau in the $T$--$\sigma$ and $T$--$\ell$
profiles, suggesting that a co-operative transition into
random coiled structure takes place. During this process, the monomers
lose contacts and the polymer is extended. As temperature is further
increased, the configuration of the polymer becomes essentially random
and there are only small changes of the total extension and the total
contact number. Similar co-operative $\beta$-sheet--coil transition
is observed when the temperature is fixed but the external force
keeps increasing. At $T=0.591 \epsilon$ Fig.~\ref{fig:fig04} indicates
the transition takes place at $f\simeq 0.5 \epsilon/a_0$.  In the 
experimental investigations of Gaub and coworkers \cite{rief1997} a
co-operative globule--extended coil  transition is observed for
the $\beta$-sheet dominated protein titin; and in a later
experiment of Fernandez and coworkers \cite{carrion-vazquez1999} 
it was suggested that
the force-induced denaturation of protein titin should follow
the same paths as the temperature-induced denaturation of this protein. 
The present theoretical work seems to confirm both of these experimental
observations: a co-operative transition from $\beta$-sheet to extended
coil will occur either by increase the environmental temperature or
by increase the mechanical force acting on the two ends of the polymer.

Conformational transitions in
single biopolymers have   attracted the interest of theorists 
for quite a long time, and the introduction of 
simple mathematical models  have brought a deep  understanding  
of the issues of the protein  $\alpha$-helix-to-coil
transition\cite{poland1970}, double-stranded DNA 
melting \cite{degennes1969}, 
$\beta$-hairpin formation in 
oligopeptides\cite{guo2000a}, and RNA secondary
structure denaturation by heat \cite{degennes1968} or 
by force \cite{zhou2001a}, etc.. 
The current work extends these efforts
to a new model system, namely, $\beta$-sheet formation in biopolymers.
It suggests that a co-operative $\beta$-sheet--coil transition
will happen at certain temperature or mechanical stress.
To make it more realistic, other 
effects such as the sequence heterogeneity and the interactions
between  adjacent layers of $\beta$-sheets \cite{lu2000}
could be considered.
On the other hand, the results presented here show that even
without these complications the structural transition
of a $\beta$-sheet is $``$two-state-like''.

Recently, the MC work in \cite{wittkop1995} suggested that pulling 
a collapsed polymer in 2D leads to a co-operative transition into 
an extended random coil, and the MC work in \cite{grassberger2001}
suggested this process actually may be a continuous transition. 
The present effort is also
complementary to these MC simulations. 

The financial support of the 
A. v. Humboldt Foundation is acknowledged.  The author is grateful to 
Professor R.\ Lipowsky for his hospitality, and to Dr.
Julian Shillcock for a careful reading of the manuscript and many good
suggestions.

\newpage
\newpage

\begin{thebibliography}{32}
\bibitem{footnote1}
The qualitative results of this paper is independent of which 
model is  used for a coil segment.  Provided no contact 
interaction between a $\beta$-sheet and a coil segment is assumed, we can
also regard a coil segment as a wormlike or 
freely-joined chain (either 2D or 3D). Here we just stick to the 
2D lattice model.
\end{thebibliography}

\newpage


\end{document}